\documentclass[slac_one]{revtex4}
\usepackage{graphicx}
\usepackage{fancyhdr}
\usepackage{xspace}
\newcommand{\DO}{D\O\xspace}
\newcommand{\MET}{\mbox{$E\kern-0.50em\raise0.10ex\hbox{/}_{T}$}}

\begin{document}

\title{Diboson Production and Couplings} 

%

\author{S. Burke on behalf of the CDF and \DO Collaborations}
\affiliation{FNAL, Batavia, IL 60510, USA}

\begin{abstract}
We present the most recent cross section measurements for $WW$ and $WZ$
production in proton-antiproton collisions with $\sqrt{s}$ = 1.96 TeV at the Fermilab
Tevatron in final states with two or three leptons, respectively,  
along with limits on $WWZ$ triple gauge couplings.  
We also present the combined search for $WW$ and $WZ$ production in the lepton, neutrino, 
and dijet final state. Finally, we present $ZZ$ cross section measurements in
both the fully leptonic and the two lepton, two neutrino final states along
with limits on anomalous couplings $ZZZ$ and $ZZ\gamma$. All results presented
are based on data collected with the Run II \DO and CDF II detectors.
\end{abstract}

\maketitle

\thispagestyle{fancy}


\section{INTRODUCTION} 

Diboson production at the Tevatron can proceed directly from quark-antiquark 
annihilation or from the boson self-interactions or Triple Gauge Couplings (TGCs) that are a 
consequence of the non-Abelian structure of the electroweak symmetry group $SU(2)_L \otimes U(1)_Y$.
Measuring diboson production provides an important test of Standard Model (SM) predictions for these
mechanisms. In particular, observing TGCs not permitted in the SM, or anomalous TGCs, 
would be a clear sign of new physics. Understanding diboson production is also critical for 
Higgs searches where dibosons are a major source of background in several important channels.  

These proceedings summarize the most recent cross section measurements by the CDF and \DO collaborations 
for $WW$, $WZ$, and $ZZ$ production, as well as limits on the corresponding TGCs. 
This includes a new 5.7$\sigma$ observation of $ZZ$ production by \DO and the world's best 
limits on anomalous $ZZZ$ and $ZZ\gamma$ production by CDF. 

\section{$WW \to l \nu l \nu$}

The simultaneous production of two $W$ bosons at the Tevatron is extremely rare with a 
predicted SM next-to-leading order (NLO) cross section of only $12.4 \pm 0.8$ pb~\cite{theory}.
\DO reported the first Tevatron observation of $WW$ production  
in events where each $W$ boson decays into an electron or a muon and a neutrino~\cite{D0:WW}. 
After selecting events with two leptons that have high transverse momenta ($p_T$) and significant 
missing transverse energy ($\MET$) from the neutrino, 25 candidate events are found  
in approximately 250 pb$^{-1}$ of data. 8.1 $\pm$ 1.0 of these events are expected to be background. The
measured cross section is $\sigma(WW)={13.8^{+4.3}_{-3.8}(stat)^{+1.2}_{-0.9}(syst)\pm 0.9(lum)}$ pb, which is 
consistent with the SM prediction. A similar measurement by CDF using approximately 825 pb$^{-1}$ of data
finds 95 candidate events, $37.8 \pm 4.8$ of which are expected to be background~\cite{CDF:WW}. 
The corresponding cross section measurement is $\sigma(WW)={13.6 \pm 2.3 (stat)\pm 1.6 (syst)\pm 1.2(lum)}$ pb, 
which is again consistent with the SM prediction.

\section{$WZ \to l \nu ll$}

$WZ$ production at the Tevatron is even more rare with a SM NLO production cross section
of $3.7 \pm 0.3$ pb~\cite{theory}. Unlike $WW$ production, which was well studied at LEP, 
$WZ$ production was first observed by the CDF collaboration using approximately 1.1 fb$^{-1}$ of data 
in the $WZ \to l \nu ll$ channel where 
$l$ is either an electron or a muon~\cite{CDF:WZobs}. \DO and CDF have both made more recent measurements
in this channel by selecting events with three high $p_T$ leptons, high $\MET$, and at least one pair of 
leptons with a combined invariant mass consistent with the mass of a $Z$ boson. Background processes  
include $Z + jets$, $Z\gamma$, $ZZ$, and $t\bar{t}$ production.
\DO finds 13 candidates in approximately 1 fb$^{-1}$ of data, 4.5 $\pm$ 0.6 of 
which are expected to be background. The corresponding measured cross section is
$\sigma(WZ)={2.7^{+1.7}_{-1.3}(stat+syst)}$ pb~\cite{D0:WZ}. CDF finds 25 candidates using 
approximately 1.9 fb$^{-1}$ of data with an expected background of $4.7 \pm 0.8$ events, and a measured 
cross section of $\sigma(WZ)={4.4^{+1.3}_{-1.0}(stat)\pm 0.2(syst)\pm 0.3(lum)}$ pb~\cite{CDF:WZ}. 
Both cross section measurements are consistent with the SM prediction.

\section{$WWZ$ Triple Gauge Couplings}

\begin{figure*}[t]
\begin{minipage}[t]{0.43\linewidth}
\includegraphics[width=0.7\linewidth]{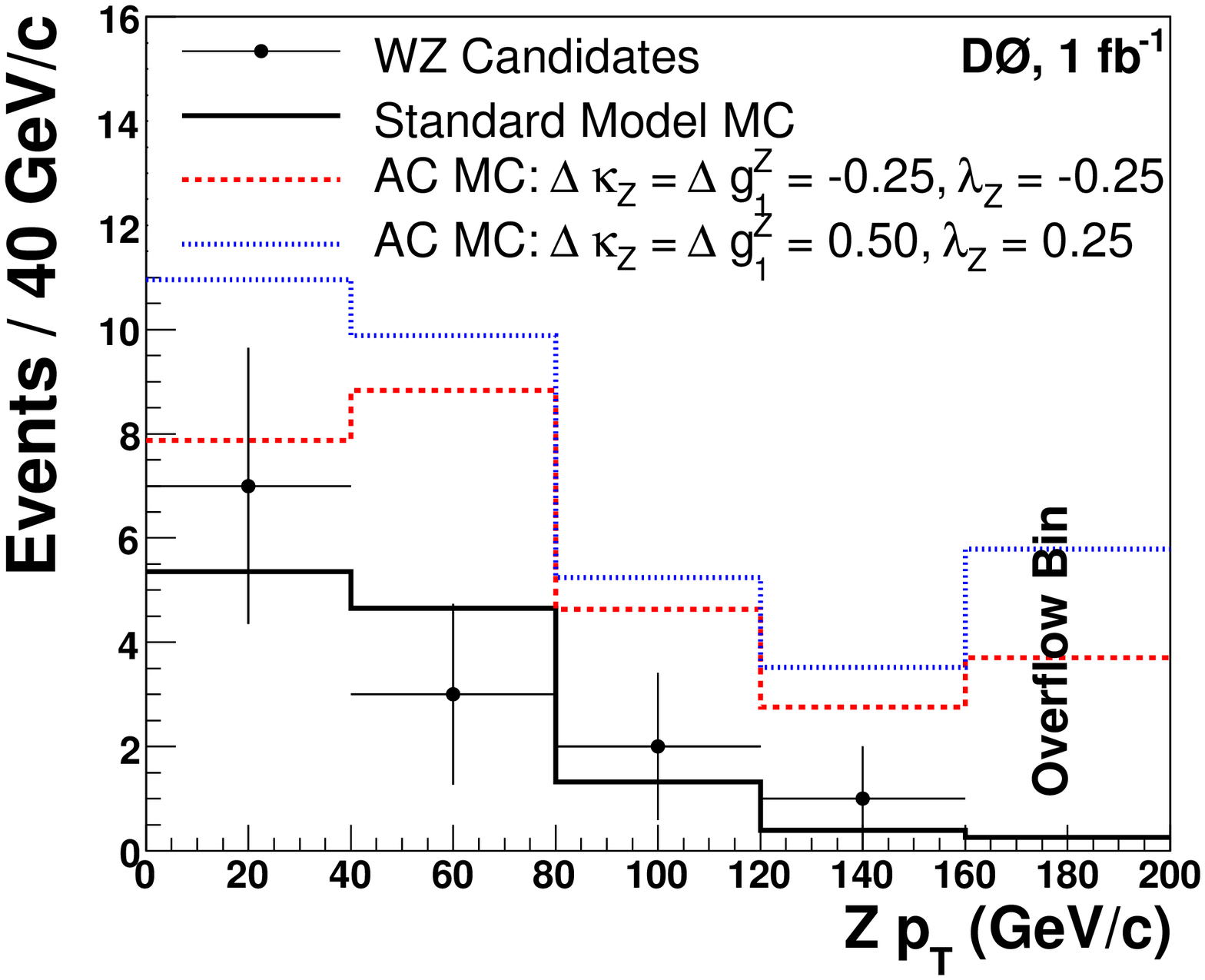}
\caption{$p_T(Z)$ distribution for $WZ$ events (\DO). The dotted lines depict possible non-SM values for $WWZ$ coupling parameters.\\~ } \label{fig:WWZ}
\end{minipage}
\hfill
\begin{minipage}[t]{0.53\linewidth}
\includegraphics[width=0.7\linewidth]{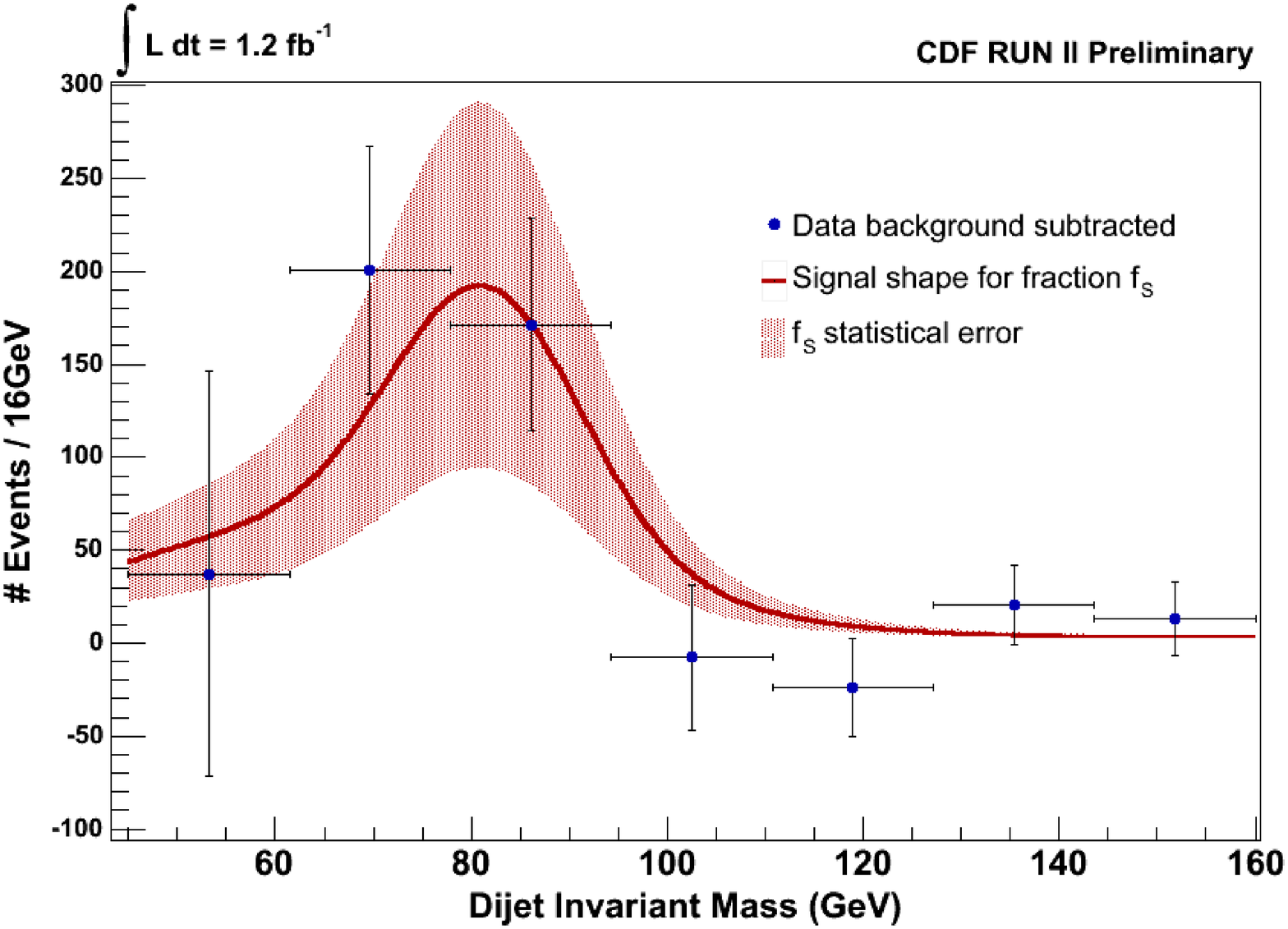}
\caption{Dijet invariant mass spectrum for $WW/WZ$ events with the background shape subtracted from data (CDF).} \label{fig:WWWZ}
\end{minipage}
\end{figure*}

In addition to measuring the cross section, both CDF and \DO set limits on $WZ$ production from 
the triple gauge coupling $WWZ$. This coupling can be described in terms of three parameters that are  
typically expressed as differences from their SM values, $\Delta g^Z_1$, $\Delta \kappa_Z$, and $\lambda_Z$. 
To set the limits, both experiments perform one- and two-dimensional
fits to their $WZ$ candidates on the $p_T$ distribution of the $Z$ boson, which is sensitive
to the coupling parameters as shown in Figure~\ref{fig:WWZ}. No excess of $WWZ$ production over the 
SM prediction is found. The one-dimensional, 95\% CL limits set by \DO assuming a form factor $\Lambda = 2.0$ TeV are: 
$-0.17 <\lambda_Z < 0.21$, $-0.14 <\Delta g^Z_1 < 0.34$, and $-0.12 <\Delta \kappa_Z = \Delta g^Z_1 < 0.29$~\cite{D0:WZ}.  
The limits set by CDF for $\Lambda = 2.0$ TeV are: $-0.13 <\lambda_Z < 0.14$, $-0.13 <\Delta g^Z_1 < 0.23$, and $-0.76 <\Delta \kappa_Z < 1.18$~\cite{CDF:WWZ}. 

\section{$WW/WZ \to l \nu jj$}

CDF has also reported a search for combined $WW$ and $WZ$ production in the partially hadronic decay
channel $WW/WZ \to l \nu jj$ using approximately 1.2 fb$^{-1}$ of data. 
Although this process has a higher branching ratio than the fully leptonic decay channel, it has not yet been observed
at a hadron collider due to the large $W+jets$ background.
Initial selection includes requiring exactly one high $p_T$ electron or muon, significant $\MET$, at least two jets, 
and a leptonic transverse mass consistent with a $W$ boson. A neural network is utilized to reduce the background. 
The signal yield, extracted from a fit to data on the dijet invariant mass 
distribution shown in Figure~\ref{fig:WWWZ}, is found to be 410 $\pm$ 212 (stat) $\pm$ 102 (syst). 
Assuming this signal (1.7$\sigma$ significance including systematics) is from WW and WZ production, the measured cross section is 
$\sigma(WW/WZ) \times BR(W \to l \nu, W/Z \to jj)={1.47 \pm 0.77 (stat)\pm 0.38 (syst)}$ pb. A 95\% CL limit 
is also set on the cross section: $\sigma(WW/WZ) \times BR(W \to l \nu, W/Z \to jj)<{2.88}$ pb~\cite{CDF:WWWZ}.

\section{$ZZ$ Production}

With a predicted NLO cross section of 1.4 $\pm$ 0.1 pb, $ZZ$ production is the most rare 
SM diboson process apart from associated Higgs production~\cite{theory}. The channels with the best sensitivity 
at the Tevatron are $ZZ \to ll \nu\nu$ and $ZZ \to llll$. 

\begin{figure*}[t]
\begin{minipage}[t]{0.48\linewidth}
\includegraphics[width=0.6\linewidth]{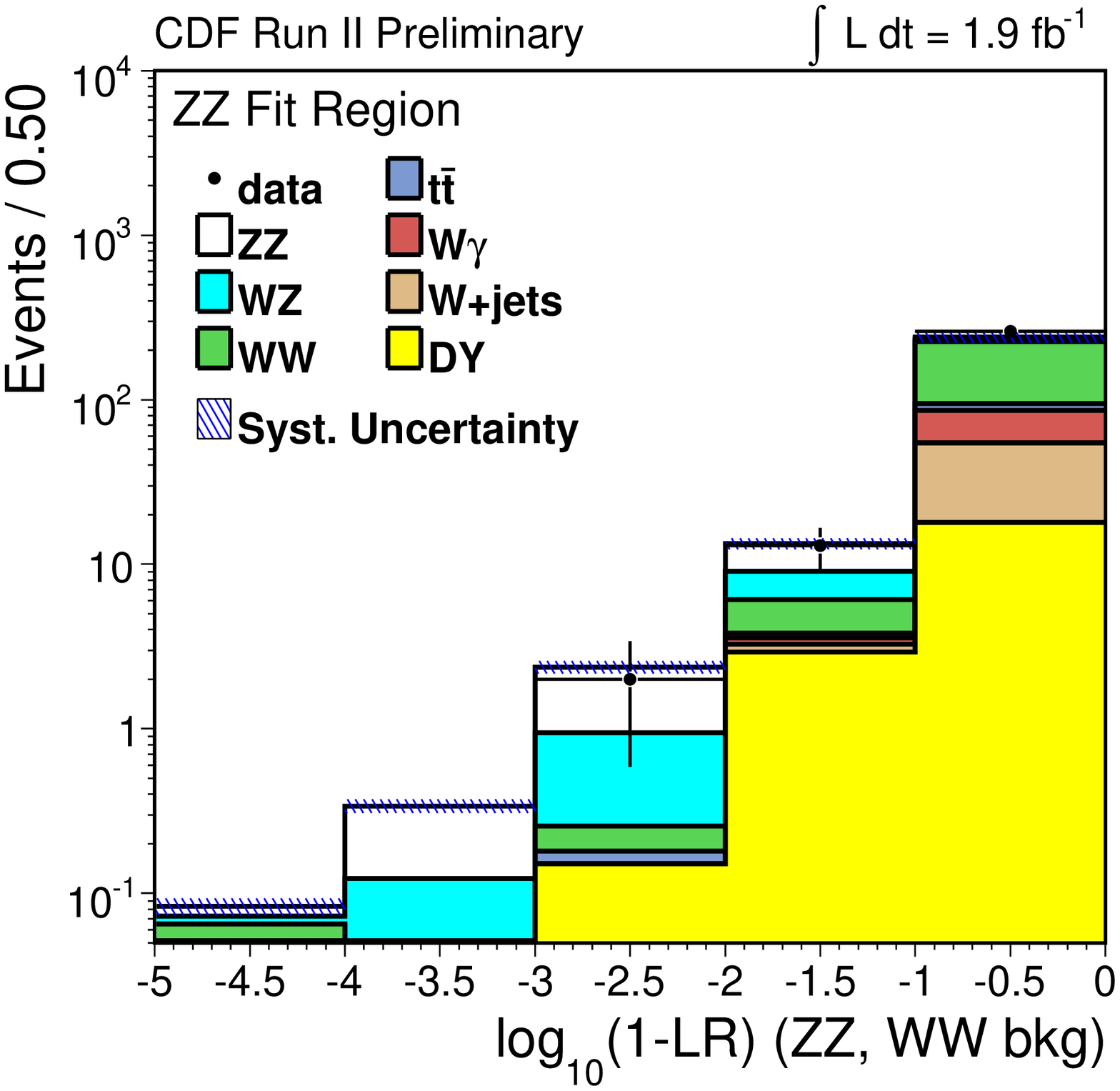}
\caption{Likelihood distribution for $ZZ \to ll \nu\nu$ (CDF). \\~} \label{fig:ZZCDF}
\end{minipage}
\hfill
\begin{minipage}[t]{0.48\linewidth}
\includegraphics[width=0.6\linewidth]{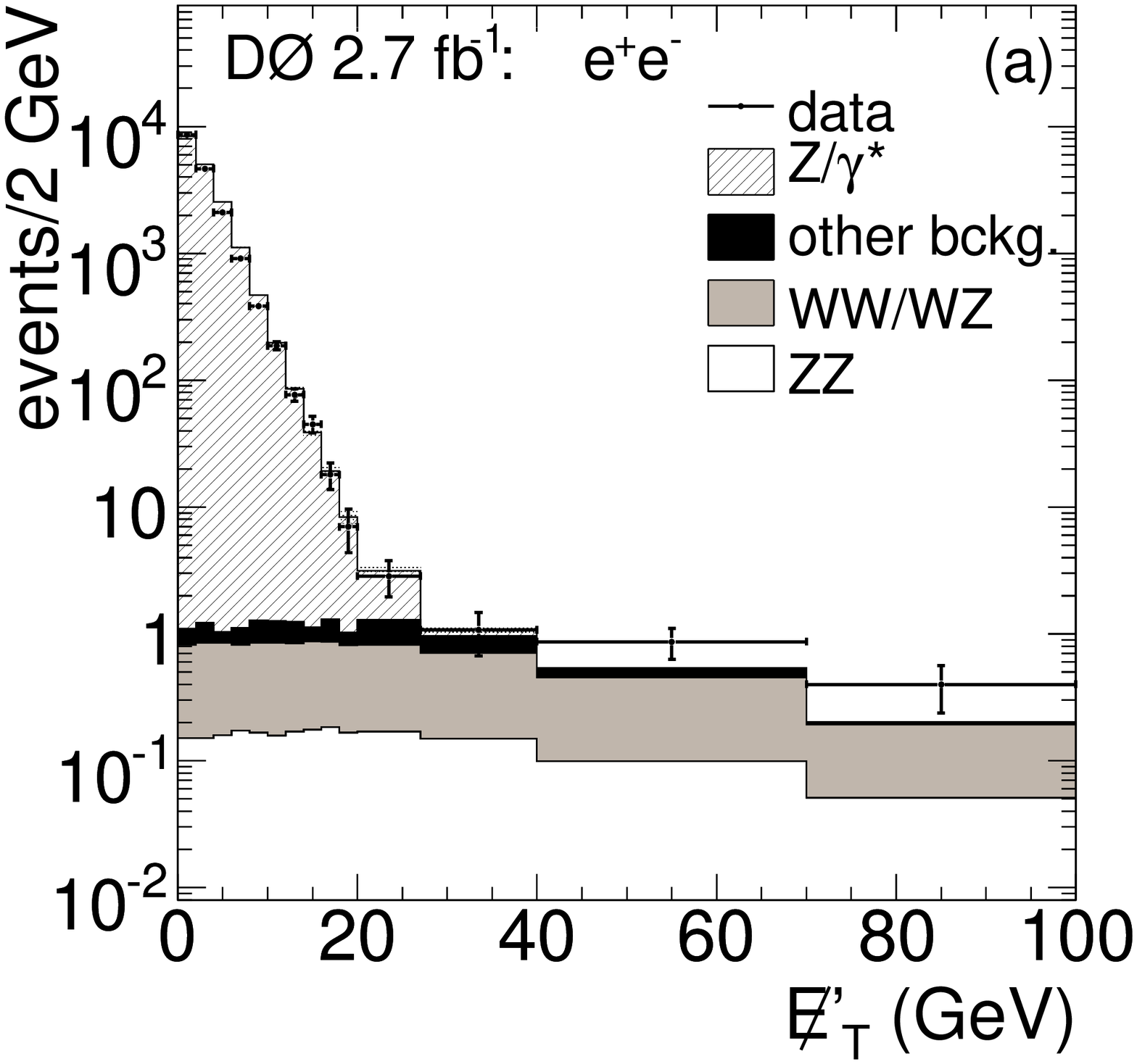}
\caption{$\MET$-like variable for $ZZ \to ee \nu\nu$ events (\DO).} \label{fig:ZZD0}
\end{minipage}
\end{figure*}

CDF published a search for $ZZ$ production using approximately 1.9 fb$^{-1}$ of data earlier this year~\cite{CDF:ZZ}.  
Initial selection in the $ZZ \to ll \nu\nu$ channel requires two high $p_T$, opposite sign electrons or muons, a 
cut on the dilepton invariant mass, and a maximum number of allowed high $p_T$ jets. 
Additional $\MET$ cuts are used to reduce the dominant $Z + jets$ background.
276 candidate events are observed, $14 \pm 2$ of which are expected to be from $ZZ$ production. 
To distinguish this signal from the large $WW$ background, a likelihood ratio, shown in Figure~\ref{fig:ZZCDF}, is constructed from 
event-by-event probability density functions that take into account the full event kinematics. The resulting signal has a significance of 1.2$\sigma$~\cite{CDF:ZZ}.
\DO's new search for $ZZ \to ll \nu\nu$ production using 2.7 fb$^{-1}$ uses similar initial selection requirements. 
However, in order to maximize discriminating power against the $Z+ jets$ background, \DO creates a novel $\MET$-like  
variable that represents the minimum possible $\MET$ in the event given the measurement uncertainties on lepton energies 
and hadronic recoil, shown in Figure~\ref{fig:ZZD0}. After cutting on this distribution, 28(15) $ee \nu \nu$ 
($\mu \mu \nu \nu$) events are found with an expected background of $15.6 \pm 0.4$ ($10.9 \pm 0.3$) events. A 
multivariate likelihood is used to distinguish $ZZ$ production from the remaining backgrounds. The resulting signal has 
a significance of 2.6$\sigma$~\cite{D0:ZZllvv}. 

Although its branching ratio is six times smaller than $ZZ \to ll \nu\nu$, the fully leptonic decay channel is the most
sensitive channel to search for $ZZ$ production at the Tevatron because few processes have 
four leptons in the final state. The dominant background, $Z+jets$, is effectively suppressed
by selecting two high $p_T$ pairs of electrons or muons that have a dilepton invariant mass consistent 
with the mass of the $Z$ boson. In order to maximize sensitivity, the published 1.9 fb$^{-1}$ CDF analysis 
separates events into higher and lower background categories based on whether the event includes an electron 
that is outside of the detector's central tracking system. One such candidate is found with an expected
background of $0.082^{+0.089}_{-0.060} (stat) \pm 0.016 (syst)$. Two candidates are found in the lower background 
category where the expected background yield is only $0.014^{+0.010}_{-0.007} (stat) \pm 0.003 (syst)$. This results in a 
signal with a significance of 4.2$\sigma$. Combined with the $ZZ \to ll\nu\nu$ channel, CDF finds a 4.4$\sigma$ signal for $ZZ$ production and 
measures a cross section of $\sigma = 1.4 ^{+0.7}_{-0.6} (stat + syst)$ pb, which is consistent with the SM 
prediction~\cite{CDF:ZZ}. \DO's new search for $ZZ \to llll$ using 1.7 fb$^{-1}$ also separates candidates into higher and lower purity samples. Seven separate categories are considered based on the number of 
electrons in the central calorimeter region. In total, one $\mu\mu\mu\mu$ candidate and 
two $eeee$ candidates are found for a total expected background of $0.14^{+0.03}_{-0.02}$ events. The observed significance of this result is 5.3$\sigma$. Combined with the 2.7 fb$^{-1}$ $ll \nu \nu$ search and an earlier $llll$ search 
on an independent 1 fb$^{-1}$ dataset~\cite{D0:ZZZ}, \DO observes $ZZ$ production with a significance of 
5.7$\sigma$~\cite{D0:ZZllll}. \DO's measured cross section, $\sigma = 1.60 \pm 0.63 (stat)^{+0.16}_{-0.17} (syst)$ pb, 
is also consistent with the SM prediction~\cite{D0:note}. 

\begin{figure*}[t]
\begin{minipage}[t]{0.34\linewidth}
\includegraphics[width=0.8\linewidth]{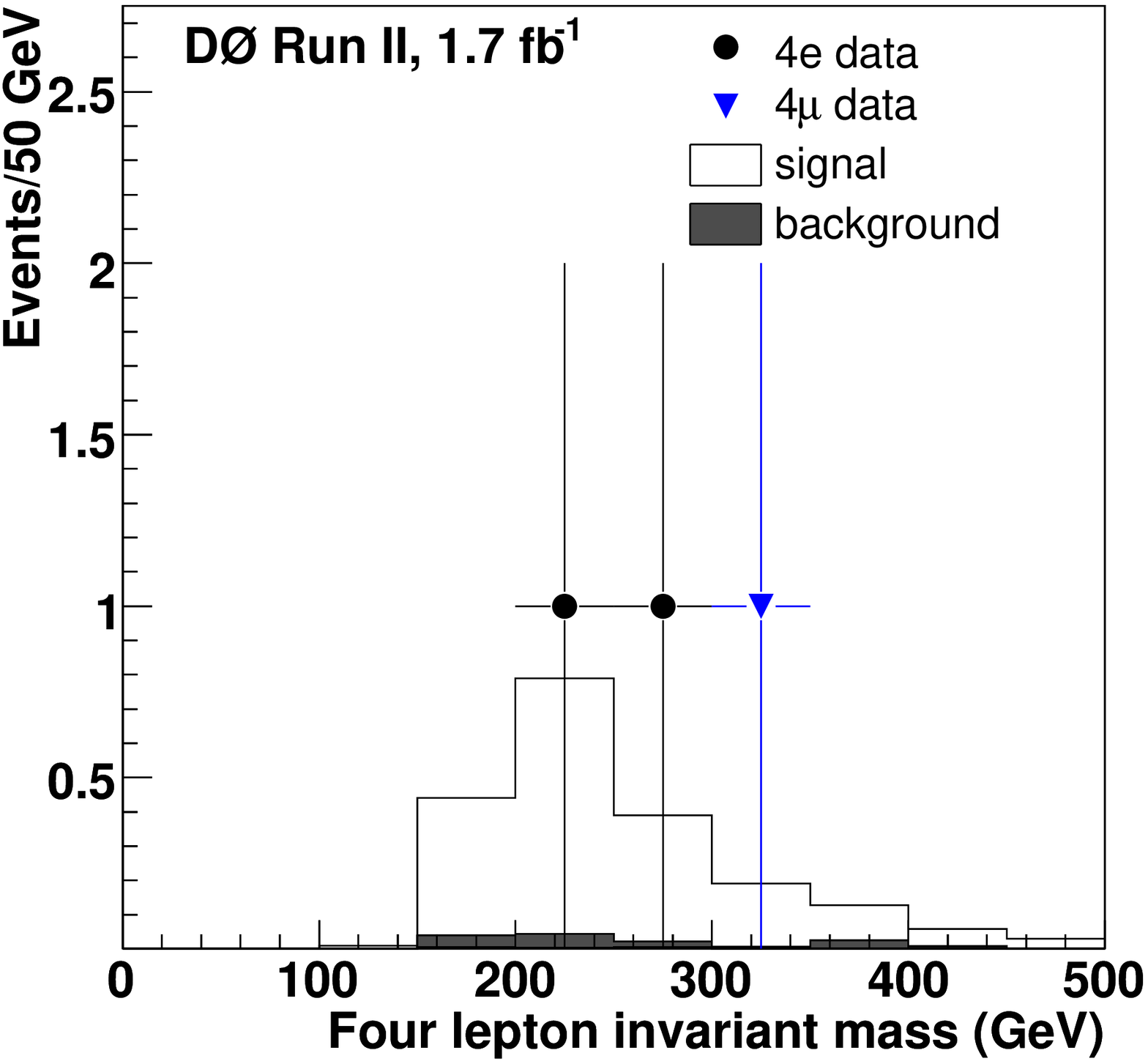}
\caption{Four-lepton invariant mass distribution for $ZZ \to llll$ (\DO).\\~}\label{fig:4lep}
\end{minipage}
\hfill
\begin{minipage}[t]{0.62\linewidth}
\includegraphics[width=0.8\linewidth]{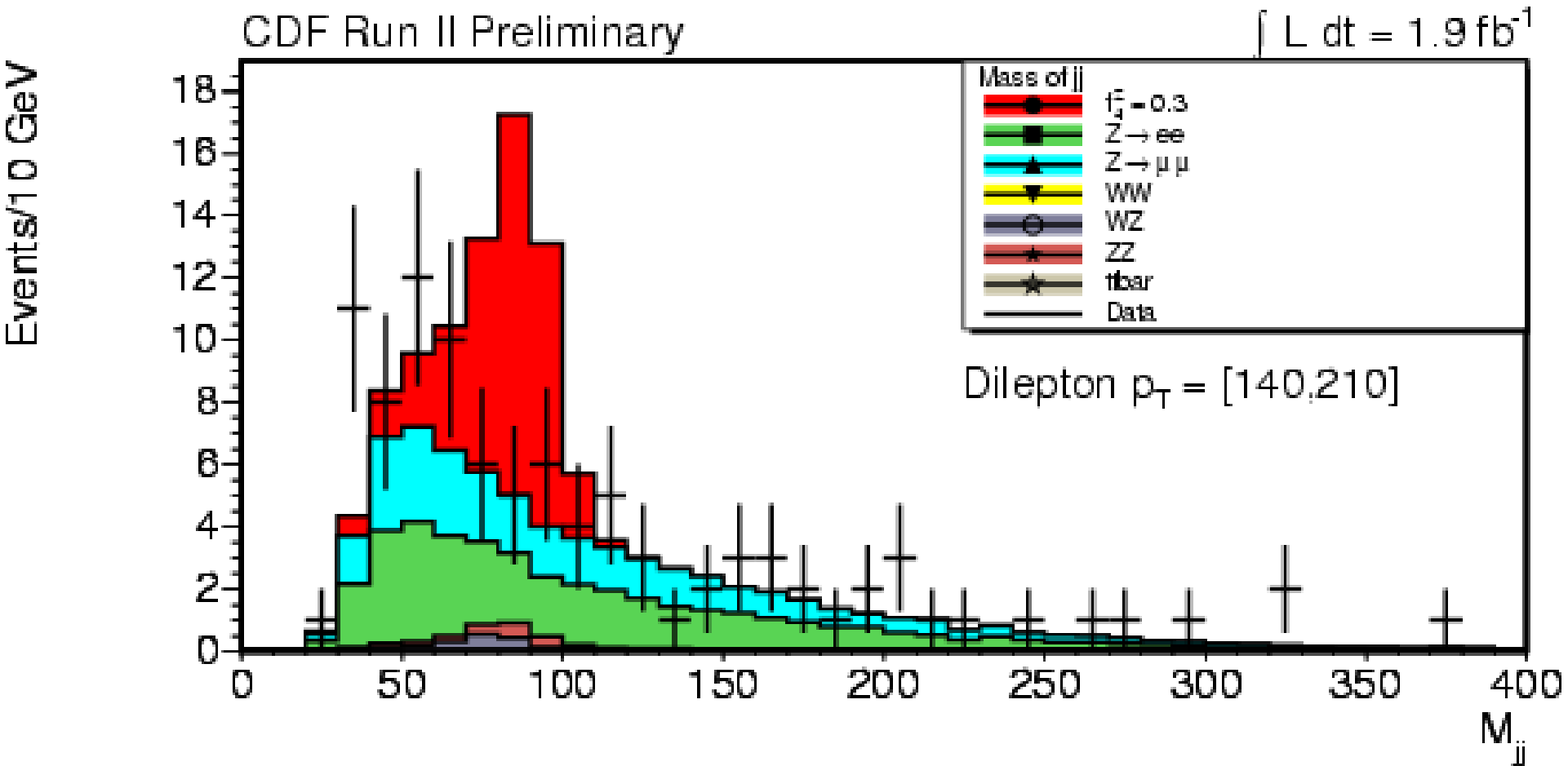}
\caption{Dijet invariant mass distribution (CDF). The red histogram depicts a possible  
value for one of the $ZZZ$ and $ZZ\gamma$ anomalous coupling parameters.}\label{fig:ZZZ}
\end{minipage}
\end{figure*}

\section{$ZZZ$ and $ZZ\gamma$ Anomalous Couplings}

The simultaneous production of two $Z$ bosons from the triple gauge couplings $ZZZ$ and $ZZ\gamma$ 
is not permitted by the SM. Both \DO and CDF have set limits on these anomalous couplings, described 
in terms of parameters $f_4^Z$, $f_4^\gamma$, $f_5^Z$, $f_5^\gamma$.  
\DO's published result based on approximately 1 fb$^{-1}$ of data set one- and two-dimensional limits on the anomalous coupling parameters using $ZZ \to llll$ candidates. The 1-dimensional $95\%$ CL limits assuming a form factor 
$\Lambda=1.2$ TeV are: $-0.28 <f_4^Z <0.28$, $-0.31 <f_5^Z <0.29$, $-0.26 < f_4^\gamma < 0.26$, 
$-0.30 < f_5^\gamma <0.28$~\cite{D0:ZZZ}. CDF has new preliminary limits based on 1.9 fb$^{-1}$ of data 
in the channel where one $Z$ boson decays to a pair of electrons or muons and the other $Z$ boson 
decays hadronically to jets. Although this
channel suffers from a large $Z+jets$ background, the background falls away at high values of the $p_T$ of the
$Z$ boson candidate formed from the two leptons. The dijet mass spectrum, shown in Figure~\ref{fig:ZZZ}, 
is fit in this high $p_T(Z)$ region to constrain potential contributions from anomalous couplings. No excess $ZZ$ production is found, 
and the world's tightest 95\% CL limits are set on the anomalous coupling parameters. For $\Lambda= 1.2$ TeV, these are: 
$-0.12<f_4^Z<0.12$, $-0.13<f_5^Z<0.12$, $-0.10<f_4^\gamma<0.10$, $-0.11 < f_5^\gamma < 0.11$~\cite{CDF:ZZZ}. 

\section{SUMMARY}
The two Tevatron experiments, CDF and \DO, have become increasingly sensitive to rare processes and have now 
measured the cross section for all SM diboson processes apart from associated Higgs production. This report 
summarized the most recent Tevatron cross section measurements for $WW$, $WZ$, and $ZZ$ production as well as 
limits on the corresponding TGCs. \DO's 5.7$\sigma$ observation of $ZZ$ production is the first 
such Tevatron measurement to cross $5\sigma$, and CDF set the world's best limits 
on anomalous $ZZZ$ and $ZZ\gamma$ production.


\begin{thebibliography}{9}
\bibitem{D0:note}
This is an updated version of the preliminary cross section measurement shown at the conference.
\end{thebibliography}
\end{document}